%% file: Frischetal_ismf_astroph.tex
\documentclass[12pt,preprint]{aastex}  

  \usepackage{times}
\def\myfunct{$F_\mathrm{i} = ~\overline{sin(\theta_\mathrm{j,i})}~ $}

\def\uncertainty{$\pm 35^\circ $}

\def\galarc{$\ell,b \sim 33^\circ ,~55^\circ $}

\def\khatb{$\hat{k}_{B,i}$}
\def\kbest{$\hat{k}_{B,best}$}
\def\microG{$\mu$G}
\def\cc{cm$^{-3}$}
\def\cmtwo{cm$^{-2}$}

\def\HI{H$^{\rm o}$}
\def\HeI{He$^{\rm o}$}

\def\ntot{$n_\mathrm{tot}$}

\def\el{\hbox{$\lambda$}}
\def\eb{\hbox{$\beta$}}
\def\elon{\hbox{$\lambda$}}
\def\elat{\hbox{$\beta$}}
\def\glon{$\ell$}
\def\glat{$b$}
\def\kms{\hbox{km s$^{-1}$}}
\def\deeg{\hbox{$^{\rm o}$}}

\begin{document}                                  


\slugcomment{Astrophysical Journal, in press }
\shorttitle{Local Interstellar Magnetic Field}
\shortauthors{Frisch et al.}

\title{Comparisons of the Interstellar Magnetic Field Directions obtained from
the IBEX Ribbon and Interstellar Polarizations}

\author{Priscilla C. Frisch}
\affil{Dept. Astronomy and Astrophysics, University of Chicago,
Chicago, IL  60637}
\email{frisch@oddjob.uchicago.edu}

\and

\author{B-G Andersson}
\affil{SOFIA, USRA}
\email{bgandersson@sofia.usra.edu}

\and

\author{Andrei Berdyugin}
\affil{Tuorla Observatory, University of Turku, Finland}
\email{andber@utu.fi}

\and
\author{Herbert O. Funsten}
\affil{Los Alamos National Laboratory, Los Alamos, NM}
\email{hfunsten@lanl.gov}

\and
\author{Mario Magalhaes} 
\affil{ Inst. de Astronomia, Geofisica, University de Sao Paulo,
  Brazil}
\email{mario@astro.iag.usp.br}

\and
\author{David J. McComas\altaffilmark{1}}
\affil{Southwest Research Institute, San Antonio, TX}
\altaffiltext{1}{also University of Texas, San Antonio, TX}
\email{DMcComas@swri.edu}

\and
\author{Vilppu Piirola}
\affil{Tuorla Observatory, University of Turku, Finland}
\email{piirola@utu.fi}

\and

\author{Nathan A. Schwadron}
\affil{Space Science Center, University of New Hampshire}
\email{nschwadron@guero.sr.unh.edu}

\and

\author{Jonathan D. Slavin}
\affil{Harvard-Smithsonian Center for Astrophysics, Harvard,
  Cambridge, MA}
\email{jslavin@cfa.harvard.edu}

\and
\author{Sloane J.  Wiktorowicz}
\affil{Astronomy Dept., University of California, Berkeley}
\email{sloane@berkeley.edu}


\begin{abstract}

Variations in the spatial configuration of the interstellar magnetic
field (ISMF) near the Sun can be constrained by comparing the ISMF
direction at the heliosphere found from the Interstellar Boundary
Explorer spacecraft (IBEX) observations of a 'Ribbon' of energetic
neutral atoms (ENAs), with the ISMF direction derived from optical
polarization data for stars within $\sim 40$ pc.  Using interstellar
polarization observations towards $\sim 30$ nearby stars within $ \sim
90^\circ$ of the heliosphere nose, we find that the best fits to the
polarization position angles are obtained for a magnetic pole directed
towards ecliptic coordinates of $\lambda,\beta \sim 263^\circ
,~37^\circ $ (or galactic coordinates of $\ell,b \sim 38^\circ
,~23^\circ $), with uncertainties of \uncertainty\ based on the broad
minimum of the best fits and the range of data quality.  This magnetic
pole is $ 33^\circ$ from the magnetic pole that is defined by the
center of the arc of the ENA Ribbon.  The IBEX ENA ribbon is seen in
sightlines that are perpendicular to the ISMF as it drapes over the
heliosphere.  The similarity of the polarization and Ribbon directions
for the local ISMF suggest that the local field is coherent over scale
sizes of tens of parsecs.  The ISMF vector direction is nearly
perpendicular to the flow of local ISM through the local standard of
rest, supporting a possible local ISM origin related to an evolved
expanding magnetized shell.  The local ISMF direction is found to have
a curious geometry with respect to the cosmic microwave background
dipole moment.
\end{abstract}


\keywords{ISM: magnetic fields, clouds, HI --- solar system: general
  --- stars: winds, outflows}


\section{Introduction} \label{sec:intro}

The recent unexpected discovery by the Interstellar Boundary Explorer
mission \citep[IBEX,][]{McComasetal:2009ssrref} of a 'Ribbon' of
energetic neutral atoms (ENA) provides the first direct evidence, and
indirect measurement, of the interstellar magnetic field embedded in
the low density partially ionized cloud surrounding the Sun
\citep{McComas:2009sci,Fuselier:2009sci,Funsten:2009sci,Schwadron:2009sci,Moebius:2009sci}.
The IBEX Ribbon is visible at 1 AU for directions where the sightline
is perpendicular to the interstellar magnetic field draping over the
heliosphere.  The center of the Ribbon arc defines a value for the
direction of the interstellar magnetic field (ISMF) in the
interstellar material (ISM) that is interacting with the heliosphere.
The ISMF directions traced by the IBEX Ribbon, and weak interstellar
polarizations observations of towards nearby stars, together provide a
unique constraint on the local ISMF direction, the outer boundary
conditions of the heliosphere, and the global magnetic turbulence in
the low density interstellar material near the Sun.  In this paper we
compare the available data on the ISMF direction from optical
polarization of local starlight, with the ISMF direction signified by
the IBEX Ribbon.

Interstellar material (ISM) sets the outer boundary conditions of the
heliosphere, including the thermal and ram pressures of the gas, and
magnetic pressures \citep[][]{Davis:1955,Parker:1961,Holzer:1989}.  The
heliosphere nose is defined by the upwind direction of the flow vector
of interstellar gas through the heliosphere, and is 15\deeg\ above the
galactic center and 5\deeg\ above the ecliptic plane.  The 0.05--6 keV
($\sim 50 - 1000$ \kms) ENAs detected by IBEX are formed in regions
where charge-exchange between interstellar neutral hydrogen and
energetic ions create inward flowing particles that reach the IBEX
detectors.  The ENAs formed inside the solar wind termination shock
propagate away from the Sun, but the subsonic solar wind and pickup
ions outside of the termination shock, with momentum components
directed back towards the Sun, can create the ENAs measured in the
inner heliosphere by IBEX
\citep[e.g.][]{Izmodenovetal:2009heliopause,Heerikhuisenetal:2008}.
The subset of ENAs that propagate back towards IBEX at 1 AU are
proving an excellent diagnostic of the heliosphere boundary
conditions.  The center of the Ribbon arc indicates an interstellar
magnetic field (ISMF) direction towards ecliptic coordinates
\el,\eb=221\deeg,39\deeg\ \citep[or in galactic coordinates towards
\glon,\glat=33\deeg,55\deeg,][]{Funsten:2009sci,Fuselier:2009sci}.
The Ribbon's arc does not follow the equator of the interstellar
magnetic field (ISMF) defined by the arc center, but rather traces the
distortion of the ISMF by the heliosphere \citep{Schwadron:2009sci}.  
For magnetic field $B$ and radial direction $R$, the length and width of the $B \cdot R$
region depend on magnetic pressure and the limit set on $B
\cdot R$, respectively.  The Ribbon latitude, in comparison to the
magnetic pole defined by the arc center, depends on the distance
beyond the heliopause of the Ribbon formation.

The ISM in the low column density sky is formed into filamentary
anisotropic structures, whether it is neutral gas seen in the 21-cm
\HI\ line, ionized gas observed in H$\alpha$ emission, or dust traced
by infrared emission or the optical polarizations of starlight.
Magnetic fields are a candidate force for controlling this observed
filamentary structure, since they generate anisotropic forces on the
ISM through pressure that acts perpendicular to the field lines, and
tension that acts parallel to the field lines \citep{Heiles:2009gan}.
 Prior to the heliospheric diagnostics of the very local ISMF, such as
the IBEX Ribbon, the only data capable of constraining the ISMF
within $\sim 50$ pc were high-sensitivity measurements of the
polarization of starlight by magnetically aligned interstellar dust
grains.  The IBEX Ribbon is a feature that can be simulated with full
heliosphere models, that simultaneously give the pressure of the ISM,
and the orientation and strength of the ISMF.  In the upwind direction
(near the galactic center), the Sun appears to be at or close to the
boundary of the surrounding cloud \citep{Frischetal:2009ibex}.  The
ENA production simulation of \citet{Heerikhuisen:2010ribbon}, which is
unconfirmed, predicts that IBEX is capable of detecting variations in
the Ribbon configuration that would be caused by variations in the
interstellar density or magnetic field, such as expected when the Sun
exits the cloud now around the heliosphere
\citep{Frischetal:2010ribbon}.  The relation between the ISMF
directions traced by the IBEX Ribbon, and the observations of the weak
interstellar polarizations of light from nearby stars, provides a
powerful tool for understanding both the distortion of the ISMF at the
heliosphere, and deviations from a homogeneous ISMF field over the
nearest tens of parsecs.

The magnetic configuration traced by the ENA ribbon, together with the
asymmetry of the heliosphere, is a function of angle between the
interstellar gas flow vector and the ISMF direction.  The flow of
interstellar \HeI\ through the heliosphere was measured by the Ulysses
probe, and showed a warm cloud $T \sim 6300 \pm340 $ K, with velocity
26.3 \kms\ and an upwind direction towards \elon,\elat$\sim
255.5^\circ,~5.1^\circ$ \citep[or
\glon,\glat=3.6\deeg,15.1\deeg,][after converting to J2000
coordinates]{Witte:2004}.  The Ribbon data therefore indicate an angle
of $\sim 46^\circ$ between the \HeI\ flow velocity and field direction.  The
gaseous component of the heliosphere boundary conditions has been
determined from models of the cloud ionization state based on
interstellar absorption lines, which indicate that the surrounding
interstellar gas is partially ionized with hydrogen $\sim 25$\%
ionized, and low density, \ntot$\sim 0.27$ \cc\ \citep[][e.g. model
26]{SlavinFrisch:2008}.  The ratio of thermal and magnetic pressures,
$\beta$, is unknown for the surrounding cloud.  If the local ISM is
near the boundaries of an evolved superbubble, which is not yet clear,
$\beta \sim 1 $ might be expected.  For the assumption of $\beta=1$,
the interstellar field strength is $\sim 2.8$ \microG\ (model 26).
For comparison, Zeeman splitting of \HI\ 21-cm lines show that
superbubble shells with filamentary structures have high ISMF
strengths and small $\beta<0.1$, although column densities are over
100 times values in the local ISM \citep{Heiles:1989shellsbeta}.  Low
$\beta$ values are also found for the Local Bubble walls foreground of
the Coalsack Nebula based on a Chandrasekhar-Fermi analysis of the
polarization component in the plane of the sky
\citep{AnderssonPotter:2006}.

The ISMF direction at the heliosphere is predicted by self-consistent
 heliosphere models
 \citep[e.g.][]{RatkiewiczBarnesSpreiter:2000,Pogorelovetal:2004,Pogorelovetal:2009,OpherRichardson:2009,Izmodenov:2009bfieldlism}
 that are constrained by the interstellar neutral and plasma density
 in the circumheliospheric ISM \citep{SlavinFrisch:2008}, the 10 AU
 difference between the Voyager 1 and Voyager 2 measurements of the
 termination shock distance \citep[of which 3--4 AU could arise from
 variable solar wind,][]{Stoneetal:2008nature}, and the $\sim 5^\circ$ offset
 between the directions of interstellar \HI\ and \HeI\ flowing into
the heliosphere \citep[][after precessing the \HeI\ and \HI\ directions into a common coordinate system]{Witte:2004,Lallementetal:2005}.
 \citet{Heerikhuisen:2010ribbon} models reproduce the IBEX Ribbon for
 an ISMF direction directed towards \el,\eb$\sim 223^\circ, 40^\circ$
 (or \glon,\glat$\sim35^\circ, 54^\circ$, hereafter \glon,\glat\ are
 galactic coordinates).  This direction is in good agreement with the
ISMF direction \glon,\glat$\sim 30^\circ \pm 5^\circ,~ 33^\circ \pm 4.3^\circ$,
for $B \sim 3.8$ \microG, from MHD models of heliospheric
asymmetries of \citet{RatkiewiczGrygorczuk:2008}.
\citet{OpherRichardson:2009} use the heliosphere asymmetries from
Voyager data and the \HI--\HeI\ offset to conclude that the ISMF
field direction is inclined by $60^\circ - 90^\circ$ with respect to
the galactic plane. Several heliosphere models suggest the very local
ISMF is directed towards the southern solar magnetic pole, such as
\citet{SwisdakOpherDrakeBibi:2010} who base their conclusion on the
assumption that the 3 kHz emissions detected by the Voyager
satellites during the early 1990's are formed by reconnection. The
general agreement between the ISMF determined from the ENA ribbon, the
Heerikhuisen et al. model, and the \citet{RatkiewiczGrygorczuk:2008} model
suggest directions for the ISMF at the heliosphere near the
center of the Ribbon arc.

The new Ribbon diagnostic of the ISMF at the heliosphere provides an
opportunity to evaluate the large-scale distortion or turbulence of
the local ISMF, e.g. over $\sim 40$ pc, through comparisons with the
polarization of starlight caused by magnetically aligned interstellar
dust grains. 
Optical polarization data show that once the curvature of spiral
arms is taken into account, the global ISMF beyond $\sim 0.3$  kpc,
is oriented towards \glon$ \sim 83^\circ$ \citep{Heiles:1996ismf}.
Faraday
rotation data show that the polarity of the global ISMF is directed towards \glon$ \sim
83^\circ$ \citep{TaylorStilSunstrum:2010}.  The Loop I magnetic
superbubble, centered about 100 pc away, is a local large distortion
of the global field with an angular diameter of $\sim 160^\circ$
\citep[e.g.][]{Heiles:1976araa,Heiles:1998whence,Wolleben:2007}.  It is
prominent in dust, gas, synchrotron emission, optical polarization,
and Faraday rotation data.  Loop I is the best candidate for the
phenomena that links the ISMF at the heliosphere with the ISMF causing
nearby star polarizations \citep{Frisch:2010s1}.  The overall geometry
of Loop I, and the 'S1' subshell, indicate that S1 has expanded to
the solar location if it is approximately spherical, and the 
ISMF direction appears consistent with nearby polarization data for a
field direction near \glon,\glat$\sim 71^\circ, ~ 48^\circ$, with
uncertainties of $\sim \pm 30^\circ$.

The ISM within $\sim 20$ pc of the Sun has very low column
densities, typically $<10^{18.5}$ \cmtwo, and therefore low
extinctions.  Pioneering efforts to measure polarized starlight caused
by magnetically aligned interstellar dust grains found that nearby
space is very empty of the ISM, except for a patch of nearby dust,
within 40 pc, primarily in the fourth Galactic quadrant between
270\deeg\ and 360\deeg, and in the southern hemisphere
\citep{Piirola:1977,Tinbergen:1982}.  This dust 'patch' has a very
local component, as the 2.5$\sigma$ detection of polarization towards
the star 36 Oph, 5 pc beyond the heliosphere nose, shows.  (It is
remarkable that many of these weak polarizations line up along the
IBEX Ribbon, see Fig. 1) The cloud giving rise to the polarization
towards 36 Oph is not the circumheliospheric ISM, since the velocities
of the two clouds differs by $\sim 2$ \kms\
\citep{Frischetal:2009ibex}, so that the polarizations in this patch
potentially trace a different ISMF direction than the Ribbon arc.  The
next set of studies of the polarization of nearby starlight was
conducted by \citet{Leroy:1993lism}, from the northern
hemisphere. Leroy made a catalog of observations of 1000 stars within
50 pc, with different sensitivity levels, and reconfirmed the
emptiness of nearby space. However, he was unable to either confirm or
disprove the existence of the polarization patch observed by
Tinbergen.  More recently, polarization observations with high
measurement accuracy have been made with the PlanetPol instrument,
primarily in the northern sky \citep{BaileyLucas:2010planetpol}.  The
Tinbergen, Piirola, and PlanetPol data are used in this study.

The goal of this study is to test the ISMF direction indicated by
the center of the arc of the IBEX Ribbon with the ISMF direction
traced by the polarization of light from nearby stars, using
mainly polarization data from the literature with a range of
sensitivity levels.  These results are preliminary, in the
sense that more and better high-sensitivity polarization data
may affect the conclusions.

\section{Local Magnetic Field Direction from Interstellar Polarizations}\label{sec:fitting}

Magnetically aligned dust grains in the ISM create a birefringent
medium with lower opacities parallel to the ISMF direction
\citep[e.g.,][]{DavisBerge:1968,Lazarian:2003}.  The position angles
of weakly polarized light, $ \lesssim 0.02$\%, from nearby stars can then be
used to trace the local ISMF direction.  A distance limit of 40 pc is
selected for polarization measurements of the local ISM, since the
``boundary'' of the Local Bubble is at about 50 pc for galactic
longitudes 300\deeg--360\deeg\ and positive latitudes
\citep[e.g.][]{Barstowetal:1997}. We assemble polarization data for
nearby stars, omitting stars with known intrinsic polarization or
circumstellar disks, and systematically evaluate the polarization
position angles for a grid of $i$ possible ISMF directions, \khatb, in
order to determine the ISMF direction (e.g. rotated coordinate system)
that provides the best alignment between polarization position angles
and the meridians of the ISMF.  The polarization data used for these
comparisons consist of three sets of archival starlight polarization
data, and new unpublished observations for three stars. The star
sample is restricted to objects within $\sim 40$ pc of the Sun.  The
combined data sets of \citet{Tinbergen:1982}\footnote{During 1973-1974
when the southern hemisphere Tinbergen data were acquired, the solar
magnetic polarity was north pole positive (A$>$0, field lines emerging
at the north pole).  The solar polarity should have no effect on the
optical polarization, as long as the grains are truly interstellar and
outside of the heliopause.}  and \citet{Piirola:1977} provide data on
$\sim 140$ stars, at $1 \sigma$ sensitivity levels of $\sim 7 \times
10^{-5} $ degree of polarization (or equivalently 0.007\% or 70 parts
per million), and in both northern and southern hemispheres.  The
strongest polarizations are seen towards the heliosphere nose region,
with strengths of $\sim 0.02$\%.  Recent PlanetPol measurements at
sensitivities of a few parts per million by
\citet{BaileyLucas:2010planetpol} add additional data in the northern
galactic and ecliptic hemispheres, where polarizations are typically
very weak ($< 0.004$\%).  Unpublished observations of several stars
are also available, including of $\beta$ Oph (HD 161096), with
polarization $P = 0.00506 \pm 0.00008$\%, and equatorial position
angle $ \theta_\mathrm{RA} = 166.2^\circ \pm 0.8^\circ$ \citep[S.
Wiktorowicz; data acquired with the POLISH
instrument,][]{Wiktorowicz:2008}, where $\theta_\mathrm{RA}$ is the
polarization position angle in equatorial coordinates.  We add
measurements of polarizations towards $\lambda$ Sgr (HD 169916) and
$\tau$ Sgr (HD 177716), that were acquired by A. Berdyugin and V. Piirola in the R-band with the DIPOL
polarimeter \citep{Piirolaetal:2005} on the KVA-60 remotely operated
telescope.  For $\lambda$ Sgr, $P=
0.036 \pm 0.007$\% and $\theta_\mathrm{RA}=109^\circ \pm 7 ^\circ $.
For $\tau$ Sgr, $P= 0.028 \pm 0.007$\% and
$\theta_\mathrm{RA}=113^\circ \pm ^\circ 7 $.  There are several stars
with data from multiple sources (e.g. $\lambda$ Sgr, $\tau$ Sgr, and
$\beta$ Oph).  In these cases all of the data were included with equal
weight in the analysis.

\begin{figure}
\plotone{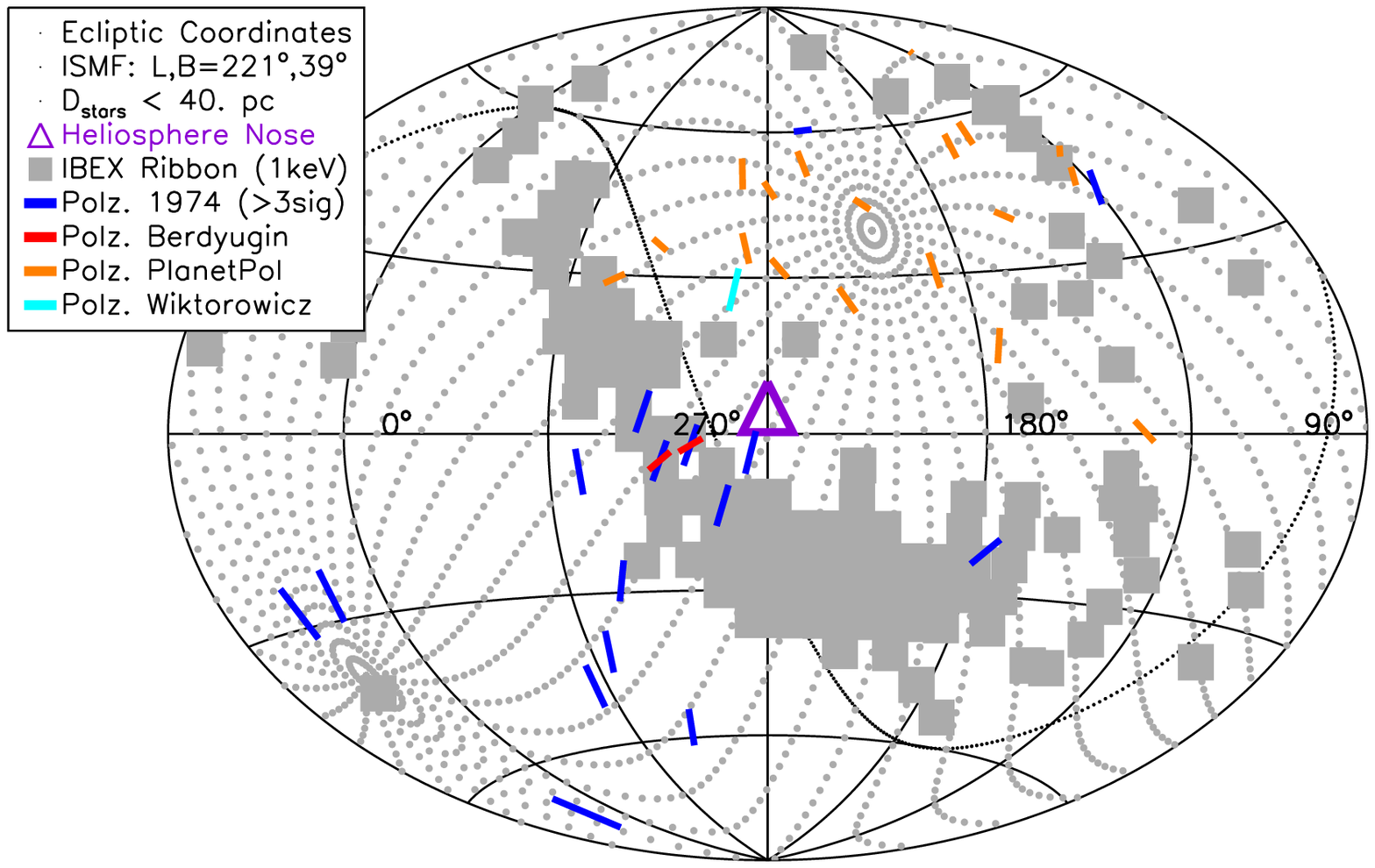}
\caption{The polarization vectors of stars within $\sim 40$ pc are
shown in the ecliptic coordinate system,
and color-coded for the data source.  The '1974' data are from
\citet[][collected in 1974]{Tinbergen:1982} and \citet{Piirola:1977}.
The plot is centered on the heliosphere nose located at ecliptic
coordinates (purple triangle) of \el=255.4\deeg, \eb=5.1\deeg, and
longitude increases towards the left in each figure.  Symbol sizes do
not indicate the strength of the polarization.  The Compton-Getting
corrected ENA fluxes at 1 keV are plotted for directions where the ENA
count rates are larger than 113 counts cm$^{-2}$ s$^{-1}$ sr$^{-1}$
keV$^{-1}$, which is 1.5 times the mean ENA flux at 1 keV as measured
by the IBEX-HI instrument \citep{McComas:2009sci}.  The dotted lines
show the ISMF determined by the center of the Ribbon arc
\citep{Funsten:2009sci}.  }\label{fig:aitoff}
\end{figure}

\begin{figure}
\plotone{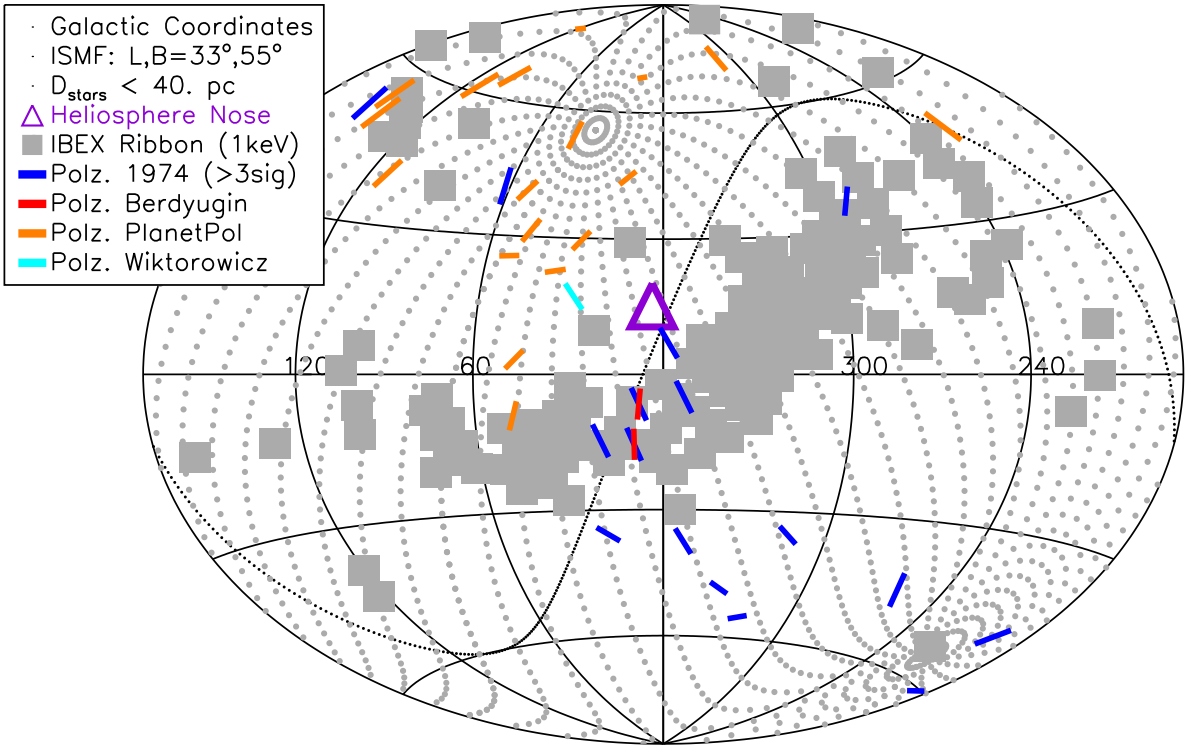}
\caption{Same as Fig. \ref{fig:aitoff}, except the figure is in galactic coordinates, and 
centered on the galactic center.  }\label{fig:aitoff2}
\end{figure}

The optical polarization position angles are plotted in galactic and
ecliptic coordinates on Aitoff projections in
Figs. \ref{fig:aitoff}, \ref{fig:aitoff2}. The plotted size of the polarization vector is
not related to the polarization strength, which spans over an order of
magnitude for these stars (see Fig. \ref{fig:ang}). The position angle
of a vector is defined with respect to the north-south meridian
passing through the position of the object, with position angle
increasing in the direction of increasing longitude.

To derive the orientation of the local ISMF we used a minimization
procedure in which we assumed that the local ISMF has a dipole
configuration, so that the variations in the orientations of the
observed polarizations are due to the location of the field poles in
the sky.  If this assumption is correct, there should be a coordinate
transformation that transforms all of the observed polarization vector
directions into vectors that are parallel to a meridian of the
transformed coordinate system.  Following \citet{Appenzeller:1968},
the position angle is transformed to coordinate system \khatb\ using
the relation:

\begin{equation}
cot(\theta_\mathrm{RA}-\theta_\mathrm{i}) = \frac{cos(b_\mathrm{i})*tan(b_\mathrm{N}) -
        cos(\ell_\mathrm{i}-\ell_\mathrm{N})*sin(b_\mathrm{i})}{sin(\ell_\mathrm{i} -\ell _\mathrm{N})}
\end{equation}
where $\theta_\mathrm{RA}$ is the position angle in the equatorial
coordinate system, and $\theta_\mathrm{i}$, $l_\mathrm{i}$, and
$b_\mathrm{i}$ are the polarization position angle and star
coordinates in the i$^{th}$ coordinate system corresponding to \khatb.
The north pole of the equatorial coordinate system is located at
$l_\mathrm{N},b_\mathrm{N}$ in the i$^{th}$ coordinate system.  Eq. 1
is used to transform between both equatorial and ecliptic or galactic
coordinates, and between equatorial coordinates and the rotated ISMF
frame \khatb.

The ISMF direction that provides the best fit to the polarization
position angles is selected by testing the polarization vector
directions against the \khatb\ grid of possible ISMF directions.  It
is assumed that the "correct" ISMF direction will be parallel to the
polarization vectors.  The grid of possible ISMF directions are spaced
by $1^\circ$ in galactic longitude, $\ell$, and latitude, $b$, over
the intervals $0^\circ < \ell_\mathrm{i} < 360^\circ$ and
$|b_\mathrm{i}| < 85^\circ$.  This comparison between polarization
position angles and the ISMF direction was restricted to the subset of
stars that have polarizations larger than $3 \sigma$, where $1 \sigma$
is the measurement uncertainty, and that are located within $\sim
90^\circ$ of the heliosphere nose.  Since the heliosphere nose is
$\sim 15^\circ$ from the galactic center, the fitting region is
essentially restricted to the first and fourth galactic quadrants
where ISM and the magnetic sky are dominated by the magnetic shell of
Loop I.  Polarization position angles for each star $j$ were first
rotated into the \khatb\ coordinate system using eq. 1, to obtain the
new position angle $\theta_\mathrm{j,i}$ relative to the new ISMF
direction corresponding to the grid point \khatb.  Polarization
position angles are degenerate with respect to the north and south
meridian directions, and the values of $\theta_\mathrm{j,i}$ were
corrected to be between 0\deeg\ and 180\deeg\ by adding or subtracting
180\deeg.

\begin{figure}
\plotone{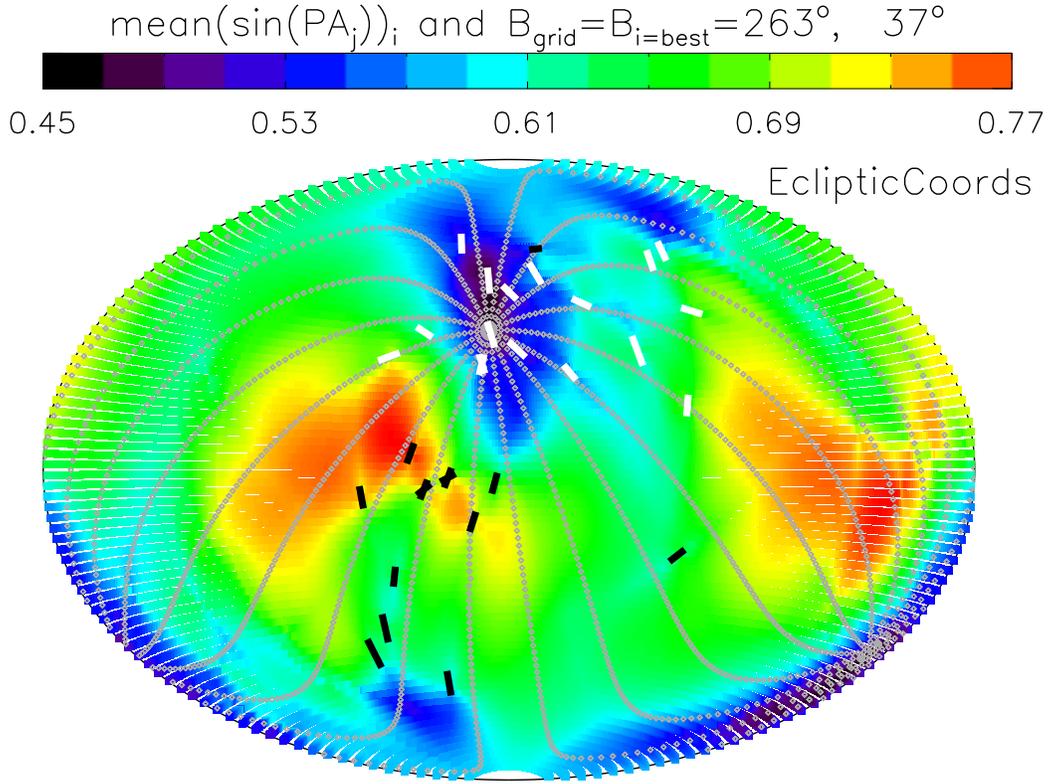}
\caption{ The value of the function \myfunct\ evaluated over a regular
grid of $i$ possible interstellar field directions (see \S
\ref{sec:fitting}).  The function is color-coded and plotted in the ecliptic coordinate
system, centered on the ecliptic nose, at $\lambda \sim 255^\circ$.  The gray dotted
grid shows the best-fit in the ecliptic 
coordinate system to the ISMF, $B_\mathrm{i=best}$, which is directed towards
$\lambda,~\beta~=~263^\circ,~37^\circ$.  Polarization vectors for the stars used in
the fit are shown as either black or white bars, for visual clarity.
}\label{fig:ismf}
\end{figure}

\begin{figure}
\plotone{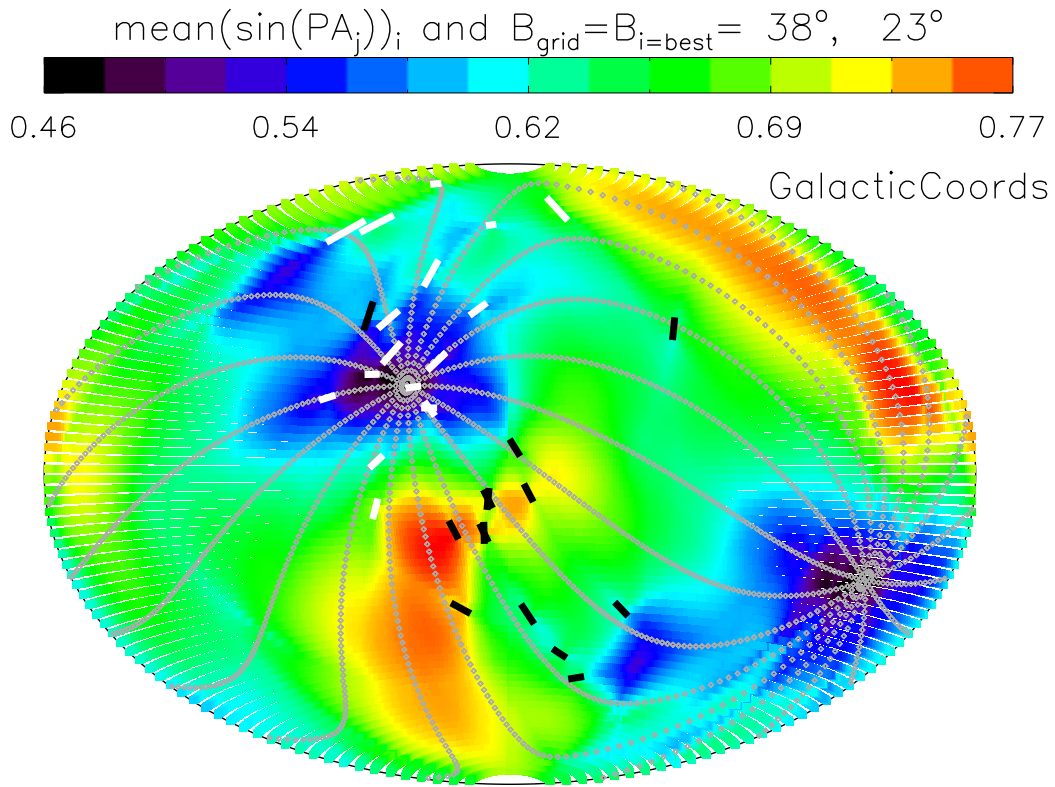}
\caption{Same as Fig. \ref{fig:ismf}, except quantities are plotted in galactic coordinates and centered on the galactic center.    The best fit to the ISMF in the galactic coordinate system,  $B_\mathrm{i=best}$, is directed towards
$\ell,~b~=~38^\circ,~23^\circ$. 
}\label{fig:ismf2}
\end{figure}

\begin{figure}
\plotone{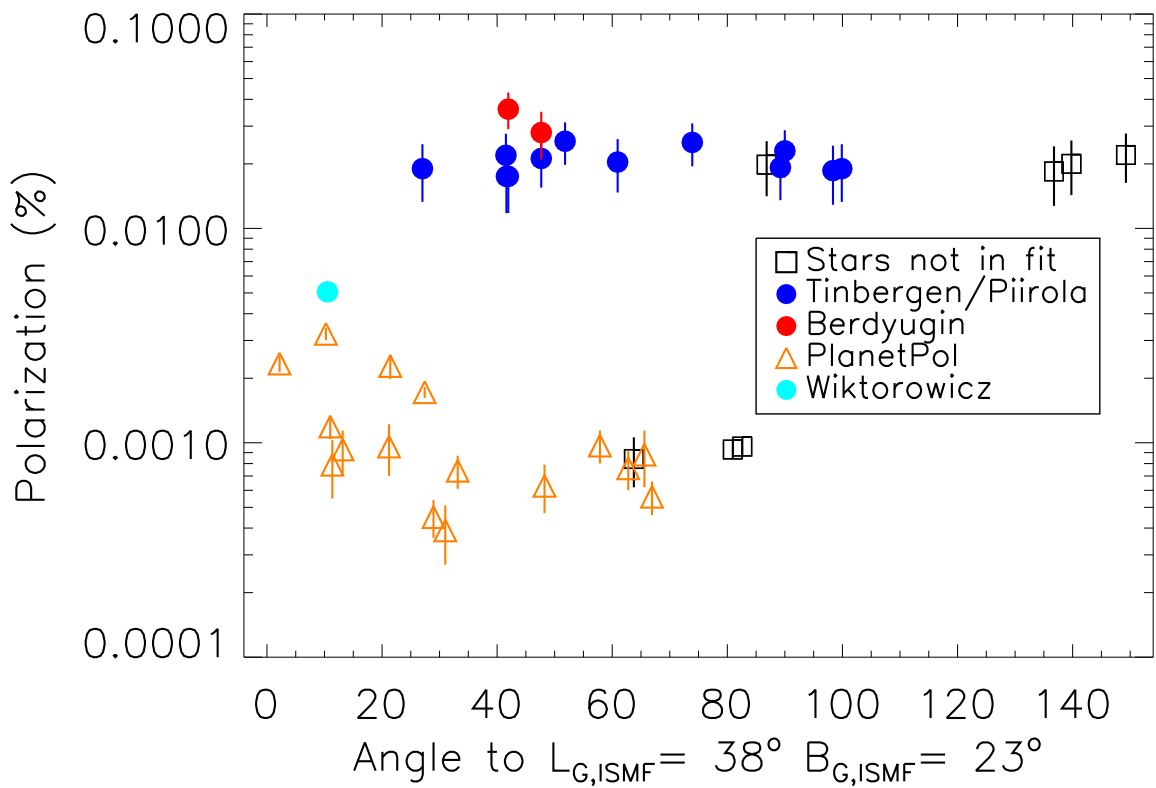}
\caption{The polarizations of stars in the combined sample (in units
of percentage of degree of polarization, ordinate) plotted against the
angular distance between the star and the north pole of the
best-fitting interstellar magnetic field, which is \glon$\sim
38^\circ$, \glat$\sim 23^\circ$.  The stars whose position angles are
used to calculate the magnetic field direction are plotted as dots or
triangles, and color coded to indicate the origin of the data.  The
stars in the full sample that have polarizations larger than the $3
\sigma$ uncertainties, but that were not used in the position angle
fit (see text), are plotted as open squares.  All stars with
polarizations less than 0.01\% have ecliptic latitudes greater than
\elat=10\deeg, while all stars with polarizations larger than 0.01\%,
except for HD 150997, are located at more negative latitudes,
\elat$<10^\circ$.  The ecliptic latitude of HD 150997 is +60\deeg, and
it is 26\deeg\ from the ISMF pole at \glon=38\deeg, \glat=23\deeg.
}\label{fig:ang}
\end{figure}

The best fitting ISMF then becomes the direction where the rotated
ISMF (the i$^{th}$ grid point) yields a minimum of some function that
describes a good match between the rotated polarization position
angles and the rotated field direction.  The function for determining
the best fitting ISMF that is adopted here is the direction where the
ISMF pole corresponds to the i$^{th}$ grid point that yields the
minimum value for the function $F_\mathrm{i}$, where \myfunct\ is the
mean sin of $j$ polarization position angles $\theta_\mathrm{j,i}$ in
the rotated coordinates.  All data points are weighted equally, rather
than by measurement accuracy, since otherwise data sampling the Ribbon
region would have lower weights since the older data (from
Tinbergen, 1982, see Fig. 1) have lower measurement accuracies.  The
position angle is the angle between the polarization vector and a
meridian, so that for a perfect fit the rotated position angles will
be 0\deeg\ or 180\deeg, and $F_\mathrm{i}=0$.  Figs. \ref{fig:ismf},\ref{fig:ismf2}
shows the color-coded map of $F_\mathrm{i}$.  The best fitting ISMF
direction, \kbest, that is determined where the minimum of
$F_\mathrm{i}=0.46$, is directed towards the ecliptic coordinates
$\lambda,\beta \sim 263^\circ ,~37^\circ $, or galactic coordinates
$\ell,b \sim 38^\circ ,~23^\circ $ (Table 1).  The broad minima shown
in Figs. \ref{fig:ismf},\ref{fig:ismf2}, together with the range of data quality,
suggest that a more accurate uncertainty for the best fit is
\uncertainty.  Figs. \ref{fig:ismf},\ref{fig:ismf2} also displays (only) the
polarization position angles that were used in the ISMF fitting
process.

\input{table1}

Several tests of the fitting process were made.  When the stellar data
set was restricted to stars within 35 pc, the best fit ISMF direction
changed by $\sim 10^\circ -20^\circ$ because four measurements near
the ecliptic equator were removed from the sample, leaving a bimodal
sample biased towards stars in the northern ecliptic hemisphere.  The
fitting function based on $F^{''}_\mathrm{i} =
~\overline{sin(\theta^2_\mathrm{j,i})}~ $ was also tried, and it gave
a best-fitting ISMF direction towards
\glon,\glat=$53^\circ,~25^\circ$, however this function overweights
position angle values near the equator of the rotated ISMF, and so
seems less suitable for this small set of data.  The fitting procedure
was repeated by varying the initial coordinate system used in the fit
(e.g. equatorial, ecliptic, or galactic), and the results agreed to
within $\sim 1^\circ$.  Another check was made by omitting the step of
converting the rotated $\theta_\mathrm{j,i}$ values to the range of
$0^\circ - 180^\circ$, yielding as expected the same result.

The polarization position angles in the rotated frame have been
constrained to be between 0\deeg\ and 180\deeg, so that the best fit
value $F_\mathrm{min} = 0.46$ corresponds to mean position angles of
27\deeg.  When the standard deviation of the position angles is
included, the best-fit mean position angle in the rotated frame is
$27^{+23}_{-19}$ degrees.  In principle the dispersion in the position
angles for the best-fitting ISMF direction could either be due to
variations in the global configuration of the nearby ISMF over
scale-lengths comparable to typical scales of energy injection, or to
small scale turbulence at scale lengths typical of the plasma and
magnetic properties of the partially ionized gas.  However the
intrinsic measurement accuracies of the data from the northern versus
southern hemisphere data sets differ substantially, so that
understanding small scale magnetic turbulence will require higher
precision data in the southern hemisphere.

For a uniform distribution of interstellar dust near the Sun, and
constant grain alignment efficiency, the polarization strengths will
increase as the angular distance between the star and the poles of the
ISMF increase, i.e. the polarizations are strongest where the
sightline is more perpendicular to the ISMF direction.  These data do
not show such an effect (Fig. \ref{fig:ang}).  Instead, stars with
ecliptic latitudes \elat$> +10^\circ$ consistently show much smaller
polarizations than stars with ecliptic latitudes below +10\deeg.  This
statement is also nearly true when galactic latitudes are used
instead.  All stars with polarizations less than 0.01\% have ecliptic
latitudes greater than \elat=10\deeg.  All stars with polarizations
larger than 0.01\%, except for HD 150997, are located at more negative
latitudes, \elat$<10^\circ$.  This effect follows from the
distribution of ISM very close to the Sun, within $\sim 15$ pc, which
has higher column densities towards negative galactic latitudes than
towards positive galactic latitudes in the galactic center hemisphere
\citep[e.g.][]{Frischetal:2009ibex}.  The ecliptic latitude of HD
150997 is +60\deeg, and it is 26\deeg\ from the ISMF pole at
\glon=38\deeg, \glat=23\deeg.  A single isolated clump of dust towards
HD 150997 is possible, or the polarization may be intrinsic to the
stellar system.  In Fig. \ref{fig:ang} the stars used in the fit are
color-coded according to the data source.  Other significant data
points, with polarizations larger than the $3 \sigma$ data
uncertainties but not used, in the fit are plotted as open squares.

Based on the above discussions we estimate uncertainties of
\uncertainty\ on the best-fit ISMF direction of $\lambda,\beta \sim
263^\circ ,~37^\circ $ in ecliptic coordinates, or $\ell,b \sim
38^\circ ,~23^\circ $ in galactic coordinates.  This direction is
$33^\circ$ from the ISMF direction at the heliosphere determined from
the arc of the IBEX Ribbon.

\section{Discussion } \label{sec:disc}

The ISMF direction of $\ell,b \sim 38^\circ ,~23^\circ $, found from
local polarization data, is directed towards the tangential region of
Loop I at the North Polar Spur (NPS), that rises 50\deeg\ vertically from the
galactic plane near \glon$\sim 30^\circ-40^\circ$.  The NPS
is a region of radio intense synchrotron emission, that \citet{Wolleben:2007}
attributes to the collision of the S1 and S2 subshells of Loop I.
Positive Faraday rotation measures
for distant pulsars and extragalactic sources towards the NPS
indicate an ISMF that is directed towards the Sun
\citep{TaylorStilSunstrum:2010}.  However at the southern galactic
latitudes for the same longitudes, where Wolleben's model suggests the
S1 subshell has expanded to the solar location, Faraday rotation measures
are negative as is consistent with an ISMF direction pointing away from
the Sun and towards the azimuthal field direction of \glon$\sim 83^\circ$. 
\citet{Salvati:2010milagrogcrs} analyzed Faraday rotation measure and dispersion 
data towards four pulsars, 150--300 pc away in the low density third galactic quadrant,
and found an ISMF directed towards \glon,\glat$=5^\circ,42^\circ$, with
strength 3.3 \microG, and with polarity directed into the
northern galactic hemisphere.  These data give the ISMF in the downwind direction,
and it is within $33^\circ$ of the best-fitting polarization ISMF direction
and within $22^\circ$ of the ISMF direction indicated
by the center of the Ribbon arc.  The good agreement between
these three independent methods of obtaining the ISMF direction
suggests the three kinds of measurements are tracing the
same ISMF, and that it is relatively smooth.

The formation mechanism for the Ribbon must be understood in order to
relate the magnetic field direction derived from the Ribbon
configuration, with the magnetic field derived from polarization data.
The 26.3 \kms\ relative motion between the heliosphere and
circumheliosphere ISM displaces the ISMF so that it drapes over the
heliosphere, and the geometry of the outer heliosphere depends on the
angle between the ISMF and interstellar flow vectors
\citep[e.g.][]{RatkiewiczBarnesSpreiter:2000,Pogorelovetal:2009,OpherRichardson:2009,Izmodenov:2009bfieldlism}.
The center of the Ribbon arc defines a magnetic field direction that
makes an angle of $\sim 46^\circ$ with the gas flow vector (Table 1).  
\citet{Heerikhuisen:2010ribbon} reproduce the
Ribbon figuration using three-dimensional MHD models of the
heliosphere plasma, coupled to interstellar neutrals described by a
kinetic distribution, and ions with a Lorentzian distribution.  In
this model, the ENAs originate upstream of the heliopause in the
region where the ISMF angle varies with the distance beyond the
heliopause. The model reproduces the location of the IBEX Ribbon quite
well for the ISMF direction of \elon,\elat$=224^\circ,~41^\circ$,
which is close to the center of the Ribbon arc.  The details of the
Heerikhuisen et al. model are not yet substantiated, because the
pitch-angle distribution of the underlying pickup ion ring-beam may
scatter over timescales shorter than the charge-exchange lifetime
\citep{McComas:2009sci,Florinskietal:2010}.  These same MHD models
also show that the Ribbon moves towards the equator of the distant
ISMF as the ENA origin pushes further upstream of the heliopause.
Several alternative scenarios have been discussed for the Ribbon
formation, including an origin in the inner heliosheath
\citep{McComas:2009sci,Schwadron:2009sci,McComasetal:2010var}.
Comparisons between magnetic field directions derived from the Ribbon
arc and polarization data will yield information on large-scale
magnetic turbulence in the solar vicinity, once the formation of the
Ribbon is fully understood.


The ISM towards the star 36 Oph (5 pc away and 10\deeg\ from the
heliosphere nose) provides insights into the ISMF and gas forming the
polarizations observed in the upwind direction.
\citet{Tinbergen:1982} observed a polarization of $\sim 0.02$\% towards 36
Oph, a strength that is unusually high compared to the mean
polarizations found over long sightlines.  A single interstellar
cloud, the 'G' cloud, is present in front of both 36 Oph (5 pc) and
the nearest star $\alpha$ Cen (1.3 pc), with a velocity that differs
by 2 \kms\ from the ISM velocity inside of the heliosphere
\citep[e.g.][and references therein]{Frischetal:2009ibex}. This
suggests that there is a single polarization screen within 1.3 pc of
the Sun in the upwind direction, and allows the possibility that the
ISMF direction derived from the polarization data is sampling a
different magnetic field than the Ribbon arc.  Heliosphere models
predict that the configuration of the IBEX Ribbon is sensitive to
20\deeg\ variations in the direction of the ISMF, so that a solar
transition between the two field directions should be readily apparent
in the configuration of the IBEX ribbon \citep{Frischetal:2010ribbon}.
The Ribbon models are not yet proven, and the uncertainties on the
ISMF direction obtained from the polarization data are large.
Nevertheless, it appears that the different ISMF directions obtained
from the center of the Ribbon arc and starlight polarizations can
provide information on the true large scale variations in the local
ISMF, and ultimately on small-scale magnetic turbulence.

The local ISMF direction derived from polarization data can be used to
test the possibility that the ISMF is perpendicular to the flow of ISM
past the Sun.  The interstellar cloud surrounding the heliosphere is
part of a cluster of local interstellar cloudlets (CLIC) that has a
mean flow velocity directed away from the center of the Loop I
magnetic superbubble \citep[e.g.][]{Frischetal:2009ibex}. For
comparing the kinematical CLIC with the geometrically defined
configuration of Loop I, the interstellar velocities are first
converted into the local standard of rest (LSR).  The standard LSR
conversion generally assumed for converting radio velocities to the
LSR is used here, e.g. a solar apex velocity of 19.5 \kms, 56\deeg,
23\deeg.  The result is an LSR upwind direction for the
CLIC that is towards
\glon,\glat=331\deeg,--5\deeg, and an LSR upwind direction for the
circumheliospheric cloud that is directed towards \glon,\glat=318\deeg,0\deeg. Both of
these directions are toward the central regions of Loop I. A value of
$\sim 71^\circ$ is found for the angle between the mean LSR flow
velocity vector of the CLIC and the best-fitting ISMF direction from
polarization data, $\ell,b \sim 38^\circ ,~23^\circ $.  The ISMF
direction is therefore nearly perpendicular to the cloud motions,
similar to what might be expected for the expansion of an evolved superbubble
shell in pressure equilibrium with ambient ISM.  For the ISM around
the heliosphere, the angle between the ISMF direction from the Ribbon
arc and the heliocentric \HeI\ flow vector is $\sim 46^\circ$.  The spatial
regions sampled by starlight polarization and CLIC interstellar absorption features are
similar, since stars within 30--40 pc of the Sun are utilized in both
comparisons.  Additional information on magnetic turbulence in the
flow of ISM past the Sun might be obtained from the
Chandrasekhar-Fermi method, by comparing the velocity dispersion of
clouds with the dispersion of polarization position angles
\citep[e.g.][]{AnderssonPotter:2006}.

There may be some contribution to the polarizations from interstellar
grains in the outer heliosheath, which will tightly follow the ISMF
that interacts with and is deflected around the heliosphere.  The
magnetic field upstream of the heliopause filters out grains with
large charge-to-mass ratios, Q/M, and gyroradii that are smaller than
the characteristic lengths between the heliosphere bow shock and
heliopause \citep[e.g.,][]{SlavinFrisch:2009sw12}.  The radii of the
magnetically excluded grains are $< 0.01 - 0.1 $ \micron, and are
comparable to the sizes of interstellar polarizing grains
\citep[e.g.][]{Mathis:1986}.  The strongest polarizations occur for
stars located on the northern edge of the IBEX Ribbon, which is
$15^\circ - 30^\circ$ above the equator of the ISMF direction towards
the arc center, corresponding to the blue points in
Fig. \ref{fig:ang}.  For truly interstellar polarization and a uniform
dust distribution, the strongest polarizations are expected at the
equator of the ISMF, so perhaps additional contributions to starlight
polarizations from nano-sized grains in the heliosheath regions are
possible.

The local ISMF direction found here has a curious coincidence with the
CMB dipole \citep[as found previously,][]{Frisch:2007cmb}.  The great
circle midway between hot and cold poles of the cosmic microwave
background dipole moment bifurcates the heliosphere nose region and is
aligned with the direction of the local ISMF direction found here, to
within the uncertainties.  The ISMF direction of $\ell,b \sim 38^\circ
,~23^\circ $ is at an angle of $90^\circ \pm 8^\circ$ from the dipoles
at \glon,\glat$=264^\circ,+48^\circ$ and $83^\circ,-48^\circ$.

\section{Conclusions} \label{sec:conclusions}

In this exploratory study we find the best-fit to the polarization
position angles towards $\sim 30$ stars within 40 pc of the Sun and $90^\circ$
of the heliosphere nose ($16^\circ$ above the galactic center),
using mainly data from the literature, is $\ell,b
\sim 38^\circ ,~23^\circ $ (or $\lambda,\beta \sim 263^\circ
,~37^\circ $, Table 1), with uncertainties of
\uncertainty\ based on the flat minimum of the best fit direction.
The ISMF direction indicated by the center of the IBEX Ribbon arc is
\galarc.  The difference between the two, $\sim 33^\circ$, is
marginally significant given the uncertainties.  The
sensitivity of the IBEX Ribbon to variations of $\le 20^\circ$ in the
ISMF direction, and the fact that the Sun is in a cloud that is
different from the nearest upwind ISM, support the possibility that
the different ISMF directions obtained from the Ribbon and
polarization data are tracking either large-scale distortion of the
magnetic field direction, or possibly small-scale magnetic turbulence.
This comparison is possible because the Ribbon is observed where the
ISMF draping over the heliosphere is perpendicular to the sightline.
Better constraints on the distortion of the ISMF in the solar vicinity
will be possible once the formation of the ENA Ribbon is understood,
and more high sensitivity polarization data towards the Ribbon are
available.

There are several implications of the ISMF direction found here.  The
ISMF vector direction is perpendicular to the bulk LSR
velocity of the cluster of local interstellar clouds flowing past
the Sun, which is consistent with an origin in an evolved expanding
magnetized superbubble shell.  The similarity of the ISMF directions found from
the ENA Ribbon and optical polarization data suggests that the nearby
ISMF is coherent over scale sizes of decades in parsecs, and that
variations in the ISMF direction due to large scale distortion or
magnetic turbulence are on the order of $ 35^\circ$ or smaller.  A
curious coincidence is that the direction of the local ISMF is within
8\deeg\ of the great circle that divides the hot and cold poles of the
cosmic microwave background dipole moment, which also passes through
the heliosphere nose.


{ \it Acknowledgements:} This research has been supported by NASA
grants NNX09AH50G and NNX08AJ33G to the University of Chicago, and by
the IBEX mission as a part of NASA's Explorer Program.  We would like
to thank Philip Lucas for sharing PlanetPol data prior to publication.

\clearpage

\clearpage

\clearpage

\clearpage

\clearpage

\end{document}

%% file: table1.tex
\begin{deluxetable}{lcc}

\tablecaption{Best-fitting Magnetic Field Pole \tablenotemark{1} }\label{tab:1} 
\tablewidth{0pt} 
\tablehead{ \colhead{Coordinate} &
\colhead{Longitude} & \colhead{Latitude} \\ \colhead{System} & & }

\startdata 
\multicolumn{2}{l}{Polarization data -- interstellar magnetic field:} & \\ 

Galactic\tablenotemark{2} & $38^\circ$ & $23^\circ $ \\ 
Ecliptic & $ 263^\circ$ & $37^\circ $ \\

\multicolumn{2}{l}{Center of Ribbon arc:} & \\
Galactic & $33^\circ$ & $55^\circ $ \\ 
Ecliptic\tablenotemark{3} & $ 221^\circ$ & $39^\circ $ \\ 
\enddata 

\tablenotetext{1}{ Galactic coordinates are
denoted by $\ell,~b$ , and ecliptic coordinates by $\lambda,~\beta$.
The estimated uncertainties on the best fit direction are
\uncertainty, based on the broad minimum for the best-fit function,
$F_\mathrm{i}$.}  

\tablenotetext{2}{This direction makes an angle of
$\sim 71^\circ$ with respect to the vector motion of the flow of
ambient local ISM past the Sun, in the LSR, which is from
\glon,\glat=331\deeg,--5\deeg\ with velocity of --19.4 \kms\
\citep{FrischSlavin:2006book}.}  

\tablenotetext{3}{This direction makes an angle of
$\sim 46^\circ$ with respect to the heliocentric
vector motion of the flow of interstellar \HeI\ into the heliosphere,
which is from
\elon,\elat$\sim 255^\circ,~ 5^\circ$ with velocity of --26.3 \kms\ \citep{Witte:2004}.}

\end{deluxetable}